\documentclass[a4paper]{jpconf}

\usepackage{amsfonts}
\usepackage{amsmath}
\usepackage{cite}
\usepackage{color}
\usepackage{xypic}
\usepackage[ps,dvips,
color,matrix,arrow]{xy}
\usepackage{mathrsfs}
\usepackage[dvips]{graphicx}
\usepackage{epstopdf}
\usepackage{graphicx}

\begin{document}

\newcommand{\scr}{\mathscr}

\def\LL{{\scr L}}
\newcommand\ZZ{{\mathbb Z}}
\def\BB{{\scr B}}

\def\fg{{\mathfrak g}}

\def\ie{{\it i.e.}}

\newcommand{\dlb}{\ensuremath{[\![}}
\newcommand{\drb}{\ensuremath{]\!]}}

\def\*{\partial}
\def\fl{\flat}
\def\sh{\sharp}

\def\tE{\widetilde E}
\def\tF{\widetilde F}
\def\tk{\widetilde k}
\def\tR{\widetilde R}
\def\ttR{\widetilde{\widetilde R}}
\def\tf{\widetilde f}

\newcommand\nn{\nonumber}
\renewcommand{\ad}{\text{ad}\,}

\frenchspacing
\parskip=4pt
\baselineskip=15pt

\title{$L_\infty$ algebras for extended geometry}
\author{Martin Cederwall and Jakob Palmkvist}
\address{Division for Theoretical Physics, Dept. of Physics}
\address{Chalmers University of Technology}
\address{SE-412 96 Gothenburg, Sweden}
\ead{martin.cederwall@chalmers.se, jakob.palmkvist@chalmers.se}
\begin{abstract}
Extended geometry provides a unified framework for double geometry,
exceptional geometry, etc., {\it i.e.}, for the geometrisations of the
string theory and M-theory dualities.
In this talk, we will explain the structure
of gauge transformations (generalised diffeomorphisms) in these
models. They are generically infinitely reducible, and arise as
derived brackets from an underlying Borcherds superalgebra or tensor
hierarchy algebra. The infinite reducibility gives rise to an $L_\infty$
structure, the brackets of which have universal expressions in terms
of the underlying superalgebra. 
\end{abstract}

\noindent The results in this talk, presented by MC at Group32,
Prague, July 9-13, 2018, are primarily based on refs.
\cite{Cederwall:2017fjm,Cederwall:2018aab}.

The motivation for the investigation lies in the dualities appearing
in string/M-theory, and the possibility to ``geometrise'' them.
The plan of the talk is to briefly mention how the dualities arise,
then display the basics of the general framework of extended geometry,
and finally discuss the gauge structure, and the appearance of an
$L_\infty$ algebra.

When M-theory is compactified on an $n$-torus, the U-duality group is
$E_{n(n)}(\ZZ)$.
The duality mixes momenta and brane windings. If this symmetry is to be “geometrised”, also diffeomorphisms and tensor gauge transformations need to be unified.

Membranes can wind on 2-cycles, 5-branes on 5-cycles.
An example: Take a 6-dimensional torus $T^6$.
There are ${6\choose2} = 15$ membrane windings $Z_{ij}$
and ${6\choose5}=6$ 5-brane windings $Z_{ijklm}$.
Collect them together with the momentum $p^i$ into the generalised
momentum
$P^M = (p^i,Z_{ij},Z_{ijklm})$.
These generalised momenta span a 27-dimensional space,
and transform under $E_{6(6)}(\ZZ)$.

The discrete duality group contains the geometric mapping class
group. It turns out that it can be “geometrised”, so that (roughly
speaking) the duality group derives from an ``extended geometry'' like the
mapping class group from geometry. 
Special cases are
double geometry (for T-duality)
\cite{Tseytlin:1990va,Siegel:1993xq,Siegel:1993bj,Hitchin:2010qz,Hull:2004in,Hull:2006va,Hull:2009mi,Hohm:2010jy,Hohm:2010pp,Jeon:2012hp,Park:2013mpa,Berman:2014jba,Cederwall:2014kxa,Cederwall:2014opa,Cederwall:2016ukd},
and exceptional geometry (for U-duality)
\cite{Hull:2007zu,Pacheco:2008ps,Hillmann:2009pp,Berman:2010is,Berman:2011pe,Coimbra:2011ky,Coimbra:2012af,Berman:2012vc,Park:2013gaj,Cederwall:2013naa,Cederwall:2013oaa,Aldazabal:2013mya,Hohm:2013pua,Blair:2013gqa,Hohm:2013vpa,Hohm:2013uia,Hohm:2014fxa,Cederwall:2015ica,Bossard:2017aae,Bossard:2018utw}.
Recently, a completely general framework has been
formulated \cite{Cederwall:2017fjm}, where any Kac--Moody group  can
be used as structure group instead of the continuous version of a duality group,
and where any integrable highest weight representation  can be used for the
generalised momenta (at
  least for highest weights with vanishing Dynkin labels
for short roots).

The gauge transformations in extended geometry --- the generalised
diffeomorphisms --- 
unify diffeomorphisms and gauge transformations for tensor fields.
Given a Kac--Moody algebra $\fg$
and a lowest weight coordinate representation $R(-\lambda)$
(we use conventions where extended tangent space vectors are in lowest
weight modules and cotangent vectors in highest weight modules;
$R(-\lambda)$ denotes the lowest weight representation with lowest
weight $-\lambda$, for dominant $\lambda$),
they are expressed in terms of a generalised Lie derivative as
$\delta_U\phi=\LL_U\phi$, for any field $\phi$ transforming
covariantly, where
\begin{align}
  \LL_UV^M=U^N\*_NV^M+Z_{PQ}{}^{MN}\*_NU^PV^Q\;.
  \label{ZGenLieDer}
\end{align}
Here the invariant tensor $Z$ has the universal expression
\cite{Bossard:2017aae,Cederwall:2017fjm}
\begin{align}
\sigma Z=-\eta_{\alpha\beta}t^\alpha\otimes t^\beta+(\lambda,\lambda)-1\;,
\end{align}
\ie, $Z_{PQ}{}^{MN}=-\eta_{\alpha\beta}(t^\alpha)_P{}^N(t^\beta)_Q{}^M
+((\lambda,\lambda)-1)\delta_P^N\delta_Q^M$, where $(t^\alpha)_M{}^N$
are representation matrices in $R(-\lambda)$, $\sigma$ is the
permutation operator and $\eta_{\alpha\beta}$ the Killing metric.

In many cases (see below), the transformations form an ``algebra''
\begin{align}
  [\LL_U,\LL_V]W=\LL_{\dlb U,V \drb}W\;,
  \label{LCommutators}
\end{align}
where the ``Courant bracket'' is
$\dlb U,V\drb=\frac12(\LL_UV-\LL_VU)$, provided that the derivatives
fulfil a {\it section constraint}.
The section constraint ensures that fields locally depend only on an
$n$-dimensional subspace of the coordinate space, on which a $GL(n)$
subgroup acts. It reads
\begin{align}
Y^{MN}{}_{PQ}\*_M\otimes\*_N=0\;,
\end{align}
where the two derivatives can act on any field (or parameter). The
tensor $Y$ is related to $Z$ as $Y=Z+1$, and can be seen (in
eq. \eqref{ZGenLieDer}) as the deviation of the generalised Lie
derivative from a Lie derivative on the extended space.
The section constraint means that any two
momenta lie in a 
linear subspace of a minimal orbit of $R(\lambda)$.

The ``algebra'' is not a Lie algebra. The consequences will be
examined below.

The closure of the generalised diffeomorphisms
as in eq. \eqref{LCommutators}  takes place only under certain conditions:
\begin{itemize}
\item[$\circ$] The algebra $\fg$ is finite-dimensional,
\item[$\circ$] $(\lambda,\theta)=1$, \ie, the highest weight $\lambda$ is a
fundamental weight dual to a simple root with Coxeter label $1$.
\end{itemize}
In other cases, so-called ancillary transformations occur. These are
restricted local $\fg$-transformations, that remove degrees of freedom
corresponding to “mixed tensors” (dual graviton, etc.)

The complete list of situations without ancillary transformations is:
\begin{itemize}
\item[({\it i})] $\fg_r=A_r$, $\lambda=\Lambda_p$, $p=1,\ldots,r$
($p$-form representations);
\item[({\it ii})] $\fg_r=B_r$, $\lambda=\Lambda_1$ (the vector representation);
\item[({\it iii})] $\fg_r=C_r$, $\lambda=\Lambda_r$ (the 
symplectic-traceless $r$-form representation);
\item[({\it iv})] $\fg_r=D_r$,
$\lambda=\Lambda_1,\Lambda_{r-1},\Lambda_r$ (the vector and spinor
representations);
\item[({\it v})] $\fg_r=E_6$, $\lambda=\Lambda_1,\Lambda_5$ (the
fundamental representations);
\item[({\it vi})] $\fg_r=E_7$, $\lambda=\Lambda_1$ (the fundamental
representation).
\end{itemize}

\begin{figure}
  \begin{center}
    {\includegraphics[scale=.4]{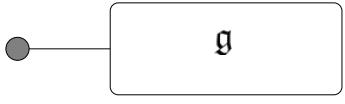}}
\end{center}
\caption{\label{BDynkin}\it Dynkin diagram of $\BB(\fg)$. The grey
  node is a fermionic null node. The single line connecting to the
  Dynkin diagram of $\fg$ can be replaced by multiple lines if
  $\lambda$ is not a fundamental weight.} 
\end{figure}

The Jacobi identity does not hold for the bracket
$\dlb\cdot,\cdot\drb$.
Instead, one typically has
\begin{align}
  \dlb U,\dlb V,W\drb\drb+\hbox{cycl.}\sim d\dlb U,V,W\drb\;,
\end{align}
where 
$\dlb U,V,W\drb\in R_2$ represents reducibility.
The derivative $d$ is linear in $\*_M$.
$R_1,R_2,\ldots$ are the positive levels (level = ghost number) of the
Borcherds superalgebra $\BB(\fg)$
\cite{Berman:2012vc,Cederwall:2015oua}
with Dynkin diagram given in Figure \ref{BDynkin}. 

This is the beginning of the $L_\infty$ structure.

What is an $L_\infty$ algebra
\cite{Lada:1992wc,Zwiebach:1992ie,Hohm:2017pnh,Roytenberg:1998vn}? 
Consider a full set of ghosts, including ghosts for ghosts, etc.
Let $C = C_1 + C_2 + C_3 + \ldots$, where the subscript indicates
ghost number.
(We will later see this as an element in $\BB_+(\fg)$, the positive
level subspace of $\BB(\fg)$.)
The Batalin--Vilkovisky (BV) action \cite{Batalin:1981jr},
restricted to ghosts, can be expanded
as
\begin{align}
S(C,C^\star)=\sum\limits_{n=1}^\infty\langle C^\star,\dlb C^N\drb\rangle\,;
\end{align}
where
\begin{align}
\dlb C^n\drb=\dlb\underbrace{C,C,\ldots,C}_n\drb
\end{align}
is the $n$-bracket.

The BV variation of $C$ is
\begin{align}
(S,C)=\sum\limits_{n=1}^\infty\dlb C^n\drb\;.
\end{align}
 In order for it to be nilpotent, $(S,(S,C)) = 0$, the brackets must
 satisfy the “generalised Jacobi identities”
 \begin{align}
  \sum\limits_{i=0}^{n-1}(i+1)\dlb C^i,\dlb C^{n-i}\drb\drb=0\;.
\end{align}
The derivative is the 1-bracket.
\begin{align}
R_1\underset{d}\leftarrow R_2\underset{d}\leftarrow R_3\underset{d}\leftarrow 
\ldots
\end{align}

\begin{figure}
  \begin{center}
    {\includegraphics[scale=1]{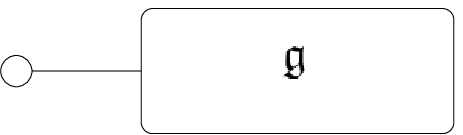}}
\end{center}
\caption{\label{ADynkin}\it The Dynkin diagrams of $\fg_{r+1}$.} 
\end{figure}

\begin{figure}
  \begin{center}
    {\includegraphics[scale=.4]{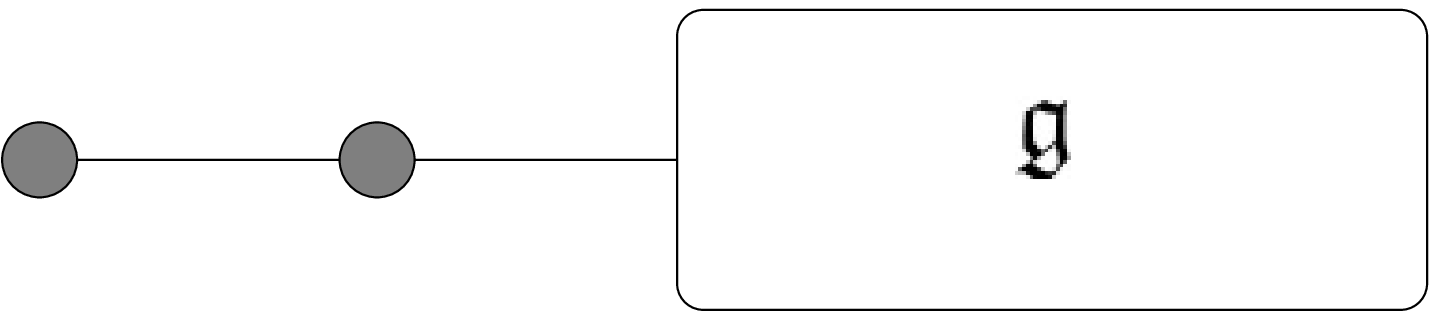}}
    $\qquad$\raise5mm\hbox{$\approx$}$\qquad$
    {\includegraphics[scale=.4]{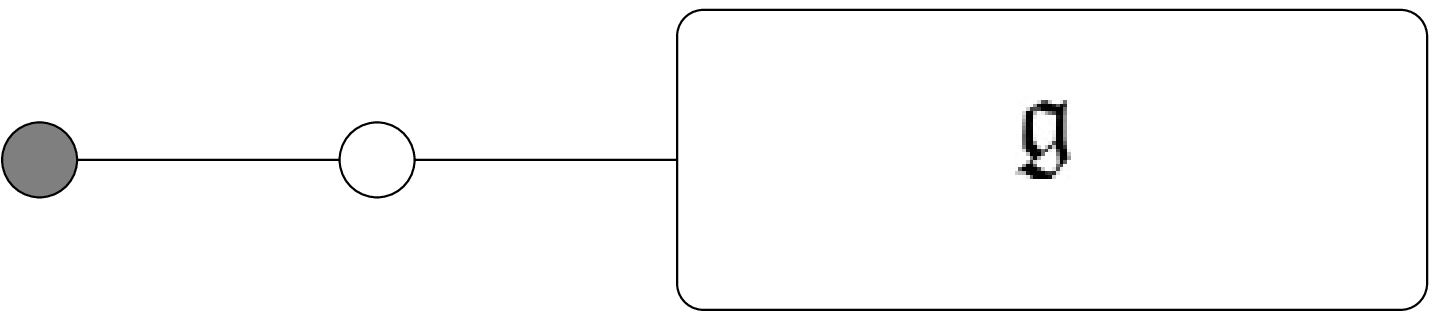}}
\end{center}
\caption{\label{CDynkin}\it Equivalent Dynkin diagrams of $\BB(\fg_{r+1})$.} 
\end{figure}

In order to be able to construct all the brackets, one needs to rely
on some “underlying” structure. It would seems like the superalgebra
$\BB(\fg)$ provides it.
However, there is no natural way of expressing the generalised Lie derivatives in terms of the (anti-)commutators of   $\BB(\fg)$.
For this one needs a further extension, $\BB(\fg_{r+1})$. The bosonic
extension $\fg_{r+1}$ has the Dynkin diagram in Figure \ref{ADynkin}
(here, like for $\BB(\fg)$, the connection of the extending node to
the Dynkin diagram of $\fg$ can consist of multiple lines). Two Dynkin
diagrams of $\BB(\fg_{r+1})$ are given in Figure \ref{CDynkin}. The
leftmost line is always single. The equivalence between the two
diagrams, expressing $\BB(\fg_{r+1})$ as an extension of $\BB(\fg)$ or
$\fg_{r+1}$, respectively, involves a fermionic Weyl reflection
\cite{Dobrev:1985qz}. 

It was noted in ref. \cite{Palmkvist:2015dea} (in that case for the
exceptional series, but straightforwardly applied to the general
situation)
that the generalised diffeomorphisms have a natural expression in
terms of the Lie super-brackets of $\BB(\fg_{r+1})$. The actual
expression is
\begin{align}
\scr L_U V = [[U,F^{\fl
      M}],\partial_{M}V^\sh]-[[\partial_{M}U^\sh,F^{\fl M}],V]\;,
\end{align}
The notation needs some explanation. We note that, in the double
grading of Table \ref{GeneralTable}, all $\fg$-modules at $p\neq0$
come in pairs. We use the nilpotent operations $\sh$ and $\fl$ to
raise and lower elements within the pairs.
$F^M$ are basis elements for $R(\lambda)$ at $p=-1$, $q=0$, and their
lowered counterpart $F^{\fl M}$ span $R(\lambda)$ at $p=-1$, $q=0$.
The two terms reproduce the two terms in eq. \eqref{ZGenLieDer}.

The doubly extended algebra is needed also in order to harbour so
called ``ancillary'' ghosts, which appear in most cases, including the
exceptional series. At some level (ghost number) the derivative fails
to be a derivation, and the generalised Lie derivative fails to be
covariant. This gives rise to the extra ancillary
ghosts. They are naturally encoded in the doubly extended algebra as
elements in $R_p$ at $q=1$. They are restricted with respect to the
section  constraint in a certain sense,
and can be constructed as a derivative together with an element in
$\tR_{p+1}$. We refer to ref. \cite{Cederwall:2018aab} for details.

\begin{table}[h]
\begin{align*}
\xymatrix@!0@C=2.2cm{
\cdots \ar@{-}[]+<1.1cm,1em>;[dddddd]+<1.1cm,-1em> \ar@{-}[]+<-0.8cm,-1em>;[rrrrrr]+<0.6cm,-1em>&
p=-1 \ar@{-}[]+<1.1cm,1em>;[dddddd]+<1.1cm,-1em>&
p=0 \ar@{-}[]+<1.1cm,1em>;[dddddd]+<1.1cm,-1em>&
p=1\ar@{-}[]+<1.1cm,1em>;[dddddd]+<1.1cm,-1em>&
p=2\ar@{-}[]+<1.1cm,1em>;[dddddd]+<1.1cm,-1em>&
p=3\ar@{-}[]+<1.1cm,1em>;[dddddd]+<1.1cm,-1em>&\cdots\\ \cdots&&&&&& *+[F-:red][red]{ n=0}\\
q=3 &&          &           &                    &     {{\ttR}}_3 \ar@{-}@[red][ur]& *+[F-:red][red]{ n=1}\\
q=2&&          &           &        {\tR_2} \ar@{-}@[red][ur]&        {\tR}_3 \oplus {{\ttR}}_3 \ar@{-}@[red][ur] & *+[F-:red][red]{ n=2 } \ar@{-}@[red][dl]\\
q=1&&{\bf 1} \ar@{-}@[red][dl]  &     R_1 \ar@{-}@[red][ur] &   R_2 \oplus \tR_2 \ar@{-}@[red][ur] &   {R}_3 \oplus {{\tR}}_3 \ar@{-}@[blue][ul] \ar@{-}@[blue][dr] & *+[F-:red][red]{ n=3 } \ar@{-}@[red][dl]\\
q=0&\overline R_1 &{\bf 1}\oplus{\bf adj}\oplus{\bf 1} \ar@{-}@[red][ur]  \ar@{-}@[red][dl]  & R_1 \ar@{-}@[blue][ul] \ar@{-}@[red][dl] \ar@{-}@[red][ur] & R_2 \ar@{-}@[blue]@[blue][ul]\ar@{-}@[red][ur]&  R_3 \ar@{-}@[blue][ul]& \cdots\\
\cdots\ar@{-}@[red][ur] &\overline R_1 & {\bf 1} & & *+[F-:blue][blue]{ \ell=1}\ar@{-}@[blue][ul] & *+[F-:blue][blue]{ \ell=2}*\frm{-}\ar@{-}@[blue][ul] & *+[F-:blue][blue]{ \ell=3}\ar@{-}@[blue][ul] 
}
\end{align*}\\
\caption{\it The general structure of the superalgebra
  ${\scr B}(\fg_{r+1})$.
  The blue lines are the $L_{\infty}$-levels, given by
  $\ell=p+q$.
  Red lines are the usual levels in the level
  decomposition of ${\scr B}(\fg_{r+1})$, and form $\fg_{r+1}$ modules.
}
\label{GeneralTable}
\end{table}

Using this Borcherds superalgebra as an underlying structure, we are
able to construct all brackets as “derived brackets” and check their
identities \cite{Cederwall:2018aab}. 

\begin{table}
\begin{align*}
  \xymatrix{
    &&&&K_{p_0}\ar[d]_\fl\ar@{<-}[r]_d
    &K_{p_0+1}\ar[d]_\fl\ar@{<-}[r]_d
    &K_{p_0+2}\ar[d]_\fl\ar@{<-}[r]_d&\cdots\\
    0\ar@{<-}[r]_d
    &C_1\ar@{<-}[r]_d
    &\cdots\ar@{<-}[r]_d
    &C_{p_0-1}\ar@{<-}[r]_d
    &C_{p_0}\ar@{<-}[r]_d
    &C_{p_0+1}\ar@{<-}[r]_d
    &C_{p_0+2}\ar@{<-}[r]_d&\cdots\\
}
\end{align*}
\caption{\it The typical structure of the action of the $1$-bracket
  between the ghost modules,
  with ancillary ghosts appearing from level $p_0\geq1$.}
\label{DerivativeActionTable}
\end{table}

The concrete expression for all brackets read:
\begin{align}
\dlb C\drb&=dC\;,\nn\\
\dlb K\drb&=dK+K^\fl\;,\nn\\
\dlb C^n\drb&=k_n\Bigl((\ad C)^{n-2}(\LL_CC+X_CC)
+\sum\limits_{i=0}^{n-3}(\ad C)^iR_C(\ad C)^{n-i-3}\LL_CC\Bigr)\;\\
\dlb C^{n-1},K\drb&=\frac{k_n}n\Bigl(
  (\ad C)^{n-2}\LL_CK+
  \sum\limits_{i=0}^{n-3}(\ad C)^i\ad K(\ad C)^{n-i-3}\LL_CC
  \Bigr)\;,\nn
\end{align}
where the coefficients have the universal model-independent expression
in terms of Bernoulli numbers
\begin{align}
  k_{n+1}=\frac{2^nB^+_n}{n!}\;,\quad n\geq1\;.
\end{align}
All non-vanishing 
brackets except the $1$-bracket contain at least one level $1$ ghost $C_1$.
No brackets contain more than one ancillary ghost.
In the expressions for the brackets, $\LL_C=\LL_{C_1}$, \ie, only the
ghost number $1$ part of $C$ enters the generalised Lie
derivative. $X$ and $R$ are ancillary contributions, and we refer to
ref. \cite{Cederwall:2018aab} for their exact expressions.

The $L_\infty$ algebra for double geometry was constructed in refs.
\cite{Deser:2016qkw,Hohm:2017pnh,Deser:2018oyg}. Then, there are no
ancillary ghosts, and the algebra stops at ghost number $2$ and a
$3$-bracket. This is because the corresponding Borcherds superalgebra
is finite-dimensional.

The Borcherds superalgebra is unable to handle
situations where ancillary transformations appear in
the commutator of two generalised Lie derivatives.
The presence of ancillary ghost at ghost number $n$ relies on the
occurrence of a $\fg$-module $\tR_p$ in $\BB(\fg_{r+1})$ (see Table
\ref{GeneralTable}). The Borcherds superalgebra never contains
$\tR_1$, which would be the signal of ancillary ghosts with ghost
number $1$, \ie, of ancillary transformations $\Sigma$ in the commutator of two
generalised  Lie derivatives,
\begin{align}
  [\LL_U,\LL_V]W=\LL_{\dlb U,V \drb}W+\Sigma_{U,V}W\;
  \label{LCommutators2}
\end{align}
The ancillary parameter $\Sigma$ is a parameter for a restricted (with
respect to the section constraint) local $\fg$-transformation, see
refs.
\cite{Hohm:2013jma,Hohm:2014fxa,Cederwall:2015ica,Bossard:2017aae}.
Then one needs a {\it tensor hierarchy algebra}
\cite{Palmkvist:2013vya,Carbone:2018njd}, a
generalisation of the Cartan-type
superalgebras $W(n)$ and $S(n)$ in Kac's classification
\cite{Kac77B}. Tensor hierarchy algebras are non-contragredient
superalgebras, and therefore {\it a priori} not defined by standard
Chevalley--Serre relations from a Dynkin diagram. In
ref. \cite{Carbone:2018njd} we presented a set of generators and
relations for the (finite-dimensional) tensor hierarchy algebras
$W(r+1)=W(A_r)$ and $S(r+1)=S(A_r)$, based on the same Dynkin diagram
as that of $\BB(A_r)$. The straightforward generalisation of these
relations seems to provide a good definition of $W(\fg)$ and $S(\fg)$ in general
(see the talk by JP at the present meeting  \cite{PalmkvistTalk}).
The superalgebra used for extended geometry with structure algebra
$\fg$ is $S(\fg_{r+1})$.
These algebras agree with the Borcherds
superalgebras at positive levels, and turn out to
``know'' when ancillary transformations appear: they contain a module
$\tR_1$ exactly in these cases (\ie, when $\fg$ is
infinite-dimensional and when $\fg$ is finite-dimensional and
$(\lambda,\theta)>1$). 
In the series of exceptional duality  symmetries, this
development is necessary starting from $E_8$.
It also seems promising for incorporating dynamical fields (vielbein,
torsion,...) in the present framework.
Work is in progress concerning the r\^ole of tensor hierarchy
algebras in extended geometry
\cite{CederwallPalmkvistTHAI,CederwallPalmkvistTHAII},
and we believe that they will encode
the information needed for the gauge structure and dynamics
also in cases with infinite-dimensional structure groups, such as
$E_9$ \cite{Bossard:2017aae,Bossard:2018utw},
$E_{10}$ and maybe $E_{11}$ \cite{Bossard:2017wxl}.

\bibliographystyle{utphysmod2}

\section*{References}


\begin{thebibliography}{10}

\bibitem{Cederwall:2017fjm}
M.~Cederwall and J.~Palmkvist,  {\em {Extended geometries}}, JHEP {\bf 02}, 071
  (2018)
[\href{http://www.arXiv.org/abs/1711.07694}{{\tt 1711.07694}}].

\bibitem{Cederwall:2018aab}
M.~Cederwall and J.~Palmkvist,  {\em {$L_\infty$ algebras for extended geometry
  from Borcherds superalgebras}},
\href{http://www.arXiv.org/abs/1804.04377}{{\tt 1804.04377}}.

\bibitem{Tseytlin:1990va}
A.~A. Tseytlin,  {\em {Duality symmetric closed string theory and interacting
  chiral scalars}}, Nucl. Phys. {\bf B350}, 395--440
(1991).

\bibitem{Siegel:1993xq}
W.~Siegel,  {\em {Two vierbein formalism for string inspired axionic gravity}},
  Phys. Rev. {\bf D47}, 5453--5459 (1993)
[\href{http://www.arXiv.org/abs/hep-th/9302036}{{\tt hep-th/9302036}}].

\bibitem{Siegel:1993bj}
W.~Siegel,  {\em {Manifest duality in low-energy superstrings}}, in {\em
  {International Conference on Strings 93 Berkeley, California, May 24-29,
  1993}}, pp.~353--363.
\newblock 1993.
\newblock
\href{http://www.arXiv.org/abs/hep-th/9308133}{{\tt hep-th/9308133}}.
\newblock

\bibitem{Hitchin:2010qz}
N.~Hitchin,  {\em {Lectures on generalized geometry}},
\href{http://www.arXiv.org/abs/1008.0973}{{\tt 1008.0973}}.

\bibitem{Hull:2004in}
C.~M. Hull,  {\em {A geometry for non-geometric string backgrounds}}, JHEP {\bf
  10}, 065 (2005)
[\href{http://www.arXiv.org/abs/hep-th/0406102}{{\tt hep-th/0406102}}].

\bibitem{Hull:2006va}
C.~M. Hull,  {\em {Doubled geometry and T-folds}}, JHEP {\bf 07}, 080 (2007)
[\href{http://www.arXiv.org/abs/hep-th/0605149}{{\tt hep-th/0605149}}].

\bibitem{Hull:2009mi}
C.~Hull and B.~Zwiebach,  {\em {Double field theory}}, JHEP {\bf 09}, 099
  (2009)
[\href{http://www.arXiv.org/abs/0904.4664}{{\tt 0904.4664}}].

\bibitem{Hohm:2010jy}
O.~Hohm, C.~Hull and B.~Zwiebach,  {\em {Background independent action for
  double field theory}}, JHEP {\bf 07}, 016 (2010)
[\href{http://www.arXiv.org/abs/1003.5027}{{\tt 1003.5027}}].

\bibitem{Hohm:2010pp}
O.~Hohm, C.~Hull and B.~Zwiebach,  {\em {Generalized metric formulation of
  double field theory}}, JHEP {\bf 08}, 008 (2010)
[\href{http://www.arXiv.org/abs/1006.4823}{{\tt 1006.4823}}].

\bibitem{Jeon:2012hp}
I.~Jeon, K.~Lee, J.-H. Park and Y.~Suh,  {\em {Stringy unification of type IIA
  and IIB supergravities under $N=2$ $D=10$ supersymmetric double field
  theory}}, Phys. Lett. {\bf B723}, 245--250 (2013)
[\href{http://www.arXiv.org/abs/1210.5078}{{\tt 1210.5078}}].

\bibitem{Park:2013mpa}
J.-H. Park,  {\em {Comments on double field theory and diffeomorphisms}}, JHEP
  {\bf 06}, 098 (2013)
[\href{http://www.arXiv.org/abs/1304.5946}{{\tt 1304.5946}}].

\bibitem{Berman:2014jba}
D.~S. Berman, M.~Cederwall and M.~J. Perry,  {\em {Global aspects of double
  geometry}}, JHEP {\bf 09}, 066 (2014)
[\href{http://www.arXiv.org/abs/1401.1311}{{\tt 1401.1311}}].

\bibitem{Cederwall:2014kxa}
M.~Cederwall,  {\em {The geometry behind double geometry}}, JHEP {\bf 09}, 070
  (2014)
[\href{http://www.arXiv.org/abs/1402.2513}{{\tt 1402.2513}}].

\bibitem{Cederwall:2014opa}
M.~Cederwall,  {\em {T-duality and non-geometric solutions from double
  geometry}}, Fortsch. Phys. {\bf 62}, 942--949 (2014)
[\href{http://www.arXiv.org/abs/1409.4463}{{\tt 1409.4463}}].

\bibitem{Cederwall:2016ukd}
M.~Cederwall,  {\em {Double supergeometry}}, JHEP {\bf 06}, 155 (2016)
[\href{http://www.arXiv.org/abs/1603.04684}{{\tt 1603.04684}}].

\bibitem{Hull:2007zu}
C.~Hull,  {\em Generalised geometry for {M}-theory}, JHEP {\bf 0707}, 079
  (2007)
[\href{http://www.arXiv.org/abs/hep-th/0701203}{{\tt hep-th/0701203}}].

\bibitem{Pacheco:2008ps}
P.~P. Pacheco and D.~Waldram,  {\em {M}-theory, exceptional generalised
  geometry and superpotentials}, JHEP {\bf 0809}, 123 (2008)
[\href{http://www.arXiv.org/abs/0804.1362}{{\tt 0804.1362}}].

\bibitem{Hillmann:2009pp}
C.~Hillmann,  {\em {$E_{7(7)}$ and $d=11$ supergravity}},
  \href{http://www.arXiv.org/abs/0902.1509}{{\tt 0902.1509}}.
PhD thesis, Humboldt-Universit\"at zu Berlin, 2008.

\bibitem{Berman:2010is}
D.~S. Berman and M.~J. Perry,  {\em Generalized geometry and {M} theory}, JHEP
  {\bf 1106}, 074 (2011)
[\href{http://www.arXiv.org/abs/1008.1763}{{\tt 1008.1763}}].

\bibitem{Berman:2011pe}
D.~S. Berman, H.~Godazgar and M.~J. Perry,  {\em {$SO(5,5)$} duality in
  {M}-theory and generalized geometry}, Phys.Lett. {\bf B700}, 65--67 (2011)
[\href{http://www.arXiv.org/abs/1103.5733}{{\tt 1103.5733}}].

\bibitem{Coimbra:2011ky}
A.~Coimbra, C.~Strickland-Constable and D.~Waldram,  {\em {$E_{d(d)} \times
  \mathbb{R}^+$ generalised geometry, connections and M theory}}, JHEP {\bf
  1402}, 054 (2014)
[\href{http://www.arXiv.org/abs/1112.3989}{{\tt 1112.3989}}].

\bibitem{Coimbra:2012af}
A.~Coimbra, C.~Strickland-Constable and D.~Waldram,  {\em Supergravity as
  generalised geometry {II}: {$E_{d(d)} \times \mathbb{R}^+$} and {M} theory},
  JHEP {\bf 1403}, 019 (2014)
[\href{http://www.arXiv.org/abs/1212.1586}{{\tt 1212.1586}}].

\bibitem{Berman:2012vc}
D.~S. Berman, M.~Cederwall, A.~Kleinschmidt and D.~C. Thompson,  {\em {The
  gauge structure of generalised diffeomorphisms}}, JHEP {\bf 01}, 064 (2013)
[\href{http://www.arXiv.org/abs/1208.5884}{{\tt 1208.5884}}].

\bibitem{Park:2013gaj}
J.-H. Park and Y.~Suh,  {\em {U}-geometry : {SL(5)}}, JHEP {\bf 04}, 147 (2013)
[\href{http://www.arXiv.org/abs/1302.1652}{{\tt 1302.1652}}].

\bibitem{Cederwall:2013naa}
M.~Cederwall, J.~Edlund and A.~Karlsson,  {\em {Exceptional geometry and tensor
  fields}}, JHEP {\bf 07}, 028 (2013)
[\href{http://www.arXiv.org/abs/1302.6736}{{\tt 1302.6736}}].

\bibitem{Cederwall:2013oaa}
M.~Cederwall,  {\em {Non-gravitational exceptional supermultiplets}}, JHEP {\bf
  07}, 025 (2013)
[\href{http://www.arXiv.org/abs/1302.6737}{{\tt 1302.6737}}].

\bibitem{Aldazabal:2013mya}
G.~Aldazabal, M.~Gra\~{n}a, D.~Marqu\'es and J.~Rosabal,  {\em {Extended
  geometry and gauged maximal supergravity}}, JHEP {\bf 1306}, 046 (2013)
[\href{http://www.arXiv.org/abs/1302.5419}{{\tt 1302.5419}}].

\bibitem{Hohm:2013pua}
O.~Hohm and H.~Samtleben,  {\em Exceptional form of ${D}=11$ supergravity},
  Phys. Rev. Lett. {\bf 111}, 231601 (2013)
[\href{http://www.arXiv.org/abs/1308.1673}{{\tt 1308.1673}}].

\bibitem{Blair:2013gqa}
C.~D. Blair, E.~Malek and J.-H. Park,  {\em {M}-theory and type {IIB} from a
  duality manifest action}, JHEP {\bf 1401}, 172 (2014)
[\href{http://www.arXiv.org/abs/1311.5109}{{\tt 1311.5109}}].

\bibitem{Hohm:2013vpa}
O.~Hohm and H.~Samtleben,  {\em Exceptional field theory {I}: {E}$_{6(6)}$
  covariant form of {M}-theory and type {IIB}}, Phys.Rev. {\bf D89}, 066016
  (2014)
[\href{http://www.arXiv.org/abs/1312.0614}{{\tt 1312.0614}}].

\bibitem{Hohm:2013uia}
O.~Hohm and H.~Samtleben,  {\em Exceptional field theory {II}: {E}$_{7(7)}$},
  Phys.Rev. {\bf D89}, 066017 (2014)
[\href{http://www.arXiv.org/abs/1312.4542}{{\tt 1312.4542}}].

\bibitem{Hohm:2014fxa}
O.~Hohm and H.~Samtleben,  {\em {Exceptional field theory. III. E$_{8(8)}$}},
  Phys. Rev. {\bf D90}, 066002 (2014)
[\href{http://www.arXiv.org/abs/1406.3348}{{\tt 1406.3348}}].

\bibitem{Cederwall:2015ica}
M.~Cederwall and J.~A. Rosabal,  {\em {E$_{8}$ geometry}}, JHEP {\bf 07}, 007
  (2015)
[\href{http://www.arXiv.org/abs/1504.04843}{{\tt 1504.04843}}].

\bibitem{Bossard:2017aae}
G.~Bossard, M.~Cederwall, A.~Kleinschmidt, J.~Palmkvist and H.~Samtleben,  {\em
  {Generalized diffeomorphisms for $E_9$}}, Phys. Rev. {\bf D96}, 106022 (2017)
[\href{http://www.arXiv.org/abs/1708.08936}{{\tt 1708.08936}}].

\bibitem{Bossard:2018utw}
G.~Bossard, F.~Ciceri, G.~Inverso, A.~Kleinschmidt and H.~Samtleben,  {\em
  {E$_9$ exceptional field theory I. The potential}},
\href{http://www.arXiv.org/abs/1811.04088}{{\tt 1811.04088}}.

\bibitem{Cederwall:2015oua}
M.~Cederwall and J.~Palmkvist,  {\em {Superalgebras, constraints and partition
  functions}}, JHEP {\bf 08}, 036 (2015)
[\href{http://www.arXiv.org/abs/1503.06215}{{\tt 1503.06215}}].

\bibitem{Lada:1992wc}
T.~Lada and J.~Stasheff,  {\em {Introduction to SH Lie algebras for
  physicists}}, Int. J. Theor. Phys. {\bf 32}, 1087--1104 (1993)
[\href{http://www.arXiv.org/abs/hep-th/9209099}{{\tt hep-th/9209099}}].

\bibitem{Zwiebach:1992ie}
B.~Zwiebach,  {\em {Closed string field theory: Quantum action and the B-V
  master equation}}, Nucl. Phys. {\bf B390}, 33--152 (1993)
[\href{http://www.arXiv.org/abs/hep-th/9206084}{{\tt hep-th/9206084}}].

\bibitem{Hohm:2017pnh}
O.~Hohm and B.~Zwiebach,  {\em {$L_{\infty}$ algebras and field theory}},
  Fortsch. Phys. {\bf 65}, 1700014 (2017)
[\href{http://www.arXiv.org/abs/1701.08824}{{\tt 1701.08824}}].

\bibitem{Roytenberg:1998vn}
D.~Roytenberg and A.~Weinstein,  {\em {Courant algebroids and strongly homotopy
  Lie algebras}},
\href{http://www.arXiv.org/abs/math/9802118}{{\tt math/9802118}}.

\bibitem{Batalin:1981jr}
I.~A. Batalin and G.~A. Vilkovisky,  {\em {Gauge algebra and quantization}},
  Phys. Lett. {\bf 102B}, 27--31
(1981).

\bibitem{Dobrev:1985qz}
V.~K. Dobrev and V.~B. Petkova,  {\em {Group theoretical approach to extended
  conformal supersymmetry: Function space realizations and invariant
  differential operators}}, Fortsch. Phys. {\bf 35}, 537
(1987).

\bibitem{Palmkvist:2015dea}
J.~Palmkvist,  {\em {Exceptional geometry and Borcherds superalgebras}}, JHEP
  {\bf 11}, 032 (2015)
[\href{http://www.arXiv.org/abs/1507.08828}{{\tt 1507.08828}}].

\bibitem{Deser:2016qkw}
A.~Deser and C.~S{\"a}mann,  {\em {Extended Riemannian geometry I: Local double
  field theory}},
\href{http://www.arXiv.org/abs/1611.02772}{{\tt 1611.02772}}.

\bibitem{Deser:2018oyg}
A.~Deser and C.~S{\"a}mann,  {\em {Derived brackets and symmetries in
  generalized geometry and double field theory}}, in {\em {17th Hellenic School
  and Workshops on Elementary Particle Physics and Gravity (CORFU2017) Corfu,
  Greece, September 2-28, 2017}}.
\newblock 2018.
\newblock
\href{http://www.arXiv.org/abs/1803.01659}{{\tt 1803.01659}}.
\newblock

\bibitem{Hohm:2013jma}
O.~Hohm and H.~Samtleben,  {\em {U-duality covariant gravity}}, JHEP {\bf 09},
  080 (2013)
[\href{http://www.arXiv.org/abs/1307.0509}{{\tt 1307.0509}}].

\bibitem{Palmkvist:2013vya}
J.~Palmkvist,  {\em {The tensor hierarchy algebra}}, J. Math. Phys. {\bf 55},
  011701 (2014)
[\href{http://www.arXiv.org/abs/1305.0018}{{\tt 1305.0018}}].

\bibitem{Carbone:2018njd}
L.~Carbone, M.~Cederwall and J.~Palmkvist,  {\em {Generators and relations for
  Lie superalgebras of Cartan type}},
\href{http://www.arXiv.org/abs/1802.05767}{{\tt 1802.05767}}.

\bibitem{Kac77B}
V.~G. Kac,  {\em {L}ie superalgebras}, Adv. Math. {\bf 26}, 8--96 (1977).

\bibitem{PalmkvistTalk}
L.~Carbone, M.~Cederwall and J.~Palmkvist,  {\em {Generators and relations for
  (generalised) Cartan superalgebras, talk by JP at Group32}},
  \href{http://www.arXiv.org/abs/yymm.nnnnn}{{\tt yymm.nnnnn}}.

\bibitem{CederwallPalmkvistTHAI}
M.~Cederwall and J.~Palmkvist,  {\em {Extended geometry and tensor hierarchy
  algebras I: Constructing the algebra from generators and relations}},
  \href{http://www.arXiv.org/abs/yymm.nnnnn}{{\tt yymm.nnnnn}}.

\bibitem{CederwallPalmkvistTHAII}
M.~Cederwall and J.~Palmkvist,  {\em {Extended geometry and tensor hierarchy
  algebras II: Gauge structure and dynamics}},
  \href{http://www.arXiv.org/abs/yymm.nnnnn}{{\tt yymm.nnnnn}}.

\bibitem{Bossard:2017wxl}
G.~Bossard, A.~Kleinschmidt, J.~Palmkvist, C.~N. Pope and E.~Sezgin,  {\em
  {Beyond $E_{11}$}}, JHEP {\bf 05}, 020 (2017)
[\href{http://www.arXiv.org/abs/1703.01305}{{\tt 1703.01305}}].

\end{thebibliography}

\providecommand{\href}[2]{#2}\begingroup\raggedright\endgroup

\end{document}